\documentclass[%
reprint,
 amsmath,amssymb,aps]{revtex4-1}

\usepackage{graphicx}
\usepackage{dcolumn}
\usepackage{amsmath,amsfonts,mathrsfs}
\usepackage{bm}
\usepackage{dodec}
\usepackage{musixtex}
\usepackage{epsfig,epstopdf}
\graphicspath{{images/}}
\providecommand{\Z}{\mathbb{Z}}

\providecommand{\VMdihedral}{\mathscr{D}}
\providecommand{\VMreste}{\mathscr{R}}

\begin{document}

\title[Sound suavity]{Measurement of the beauty of periodic noises}
\author{Vincent M. MANET}
\affiliation{Marmonier, 35, rue du dauphin\'e, 69960 Corbas, France}
\email{vincent.manet@marmonier.com}

\date{\today}

\begin{abstract}
In this article indicators to describe the ``beauty''
of noises
are proposed.
Rhythmic, tonal and harmonic suavity are introduced.
They give a characterization of a noise in terms of rhythmic regularity (rhythmic
suavity), of auditory pleasure of the ``chords''  constituting
the signal (tonal suavity) and of the transition between the chords (harmonic suavity).
These indicators have been developed for periodic noises typically
issued from rotating machines such as engines, compressors... 
and are now used by our industrial customers since two years.
\end{abstract}

\maketitle

\section{Introduction}
In the industry, we often have to improve a noisy environment.
The first way, after having identified the sources, consists of lowering
the noise level arriving to the user.
Hovewer nowadays, this is not sufficient.
Conforming to acoustic standards, even drastic ones, does not imply 
customer's satisfaction.
Indeed, to endure a noise during hours, even with a moderate intensity, is
only possible if this noise has a certain quality, a certain ``beauty''.

\medskip
Moreover, we face more and more emergency situations, which means that
solutions have to be developed and installed very quickly.
We cannot use sequential trials nor ask customers about their feelings.
We have to directly furnish the ``right'' solution in terms of noise level
as well as sound comfort (sound beauty).
This is why it is necessary to be able to characterize this beauty in some sense.

\medskip
In this article, we present several indicators, each one describing the
sound beauty according to a point of view. These indicators, defined as
values between 0 (bad) and 1 (perfect), can then be combined into 
a global indicator of the considered noise.

\medskip
To approach the subject of the beauty of a noise is fundamentally 
different from that for example the music. 
This is why no reference is made ​​to works of classical psychoacoustics.
We want to emphasize that we only deal with noises, and more precisely 
with ``neutral'' noises: a neutral noise is a noise to which no cultural nor emotional 
component is attached. 
For example, the noise of a car engine is not neutral, because its perception is 
clearly related to cultural and emotional factors.
But noises of an air conditioner, of an industrial compressor or of the engine 
of a mechanical tool are neutral.

\section{Stage 1 (optional): determination of ``cycles'' in the acoustic signal}

In music (played by musicians, not generated by computers), time between two successive
pulses cannot generally be less than 20~ms (even if smaller values can be found in more contemporary 
musical forms).
But Green\cite{bib-green} and Roads\cite{bib-roads} showed that events separated by only 1~ms 
can be detected by human ears.

Therefore, we do not impose any minimum duration of cycles that we study. 
This is especially true for periodic signals, where even a very small part of the signal is important 
because it is repeated again and again.

\medskip
Since we study periodic signals, we must be able to isolate a signal cycle to 
examine it in detail.

\begin{figure*}
   \centerline{\epsfig{figure=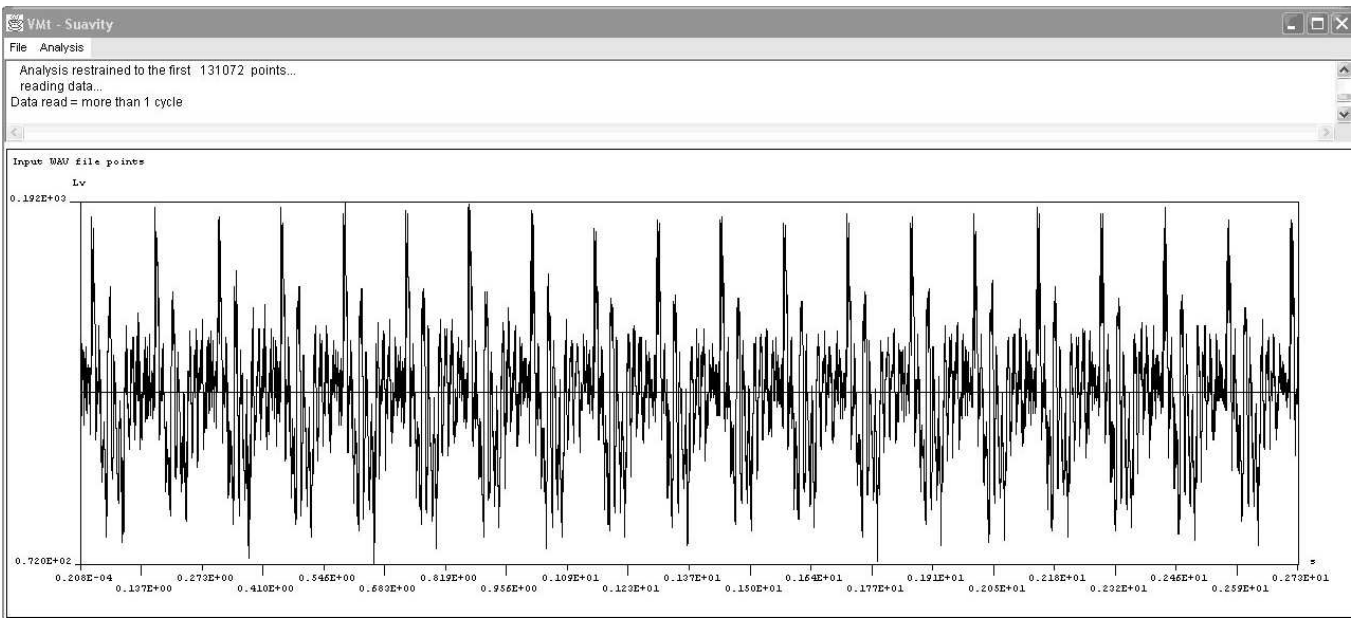,width=\textwidth}}
   \caption{\label{fig:typique}Typical temporal recording of a periodic signal}
\end{figure*}

Fig.~\ref{fig:typique} presents a typical temporal signal of a periodic phenomenon.
Studied signals (cyclic noise emitted by rotating machines) have the distinction of having 
an emerging component at each cycle.
The algorithm exploits this property: a first approximation of the cycle time is obtained by 
studying the lowest frequencies of the spectrum. This approximation is then refined by 
the study of the temporal signal.
This gives excellent results, as illustrated by industrial examples presented at the end of the article.

If this algorithm is not adapted to the kind of studied signal, we suppose that a cycle is extracted using 
another method (or manually), and in following stages, we consider only one cycle of the signal.

\section{Stage 2: Rhythmic suavity}

This indicator represents the rhythmic regularity of an acoustic signal, and
is based on a temporal analysis of the signal.
For example, an engine can rotate in a uniform way
or not (without taking into account the sound it produces).
This feeling of regularity or of irregularity is already a part (a component)
of the beauty of the acoustic signal.
In order to define it, we propose to find the pulses within the considered
acoustic signal, i.e. to find where acoustic events occur.
The rhythmic suavity will be defined as the number of acoustic events divided
by the total number of pulsations within the whole acoustic signal.

\medskip
Software having a ``beat finder'' function often use a dynamical approach (in a musical
sense) in order to represent the notion of beginning (or ending) of a note.
Pulses are located where voices come into play: 
a slope greater than a given threshold is hunted in the temporal signal, as for 
example in Audacity\cite{bib-Audacity} software.
This notion can also be studied in a spectral way, as for example, in 
AudioSculpt\cite{bib-Audiosculpt} software.

For short periodic signals (which are our concern), methods based on the spectrum 
derivative or variation are not adapted because notions such as beginning or ending of 
notes do not exist: the signal is too smooth (compared to music), and to lower the
threshold leads to very unstable results.
Smoothing methods (as windowed average) applied to the signal leads to the same
kind of threshold problems.

\begin{figure}
             \centerline{\epsfig{figure=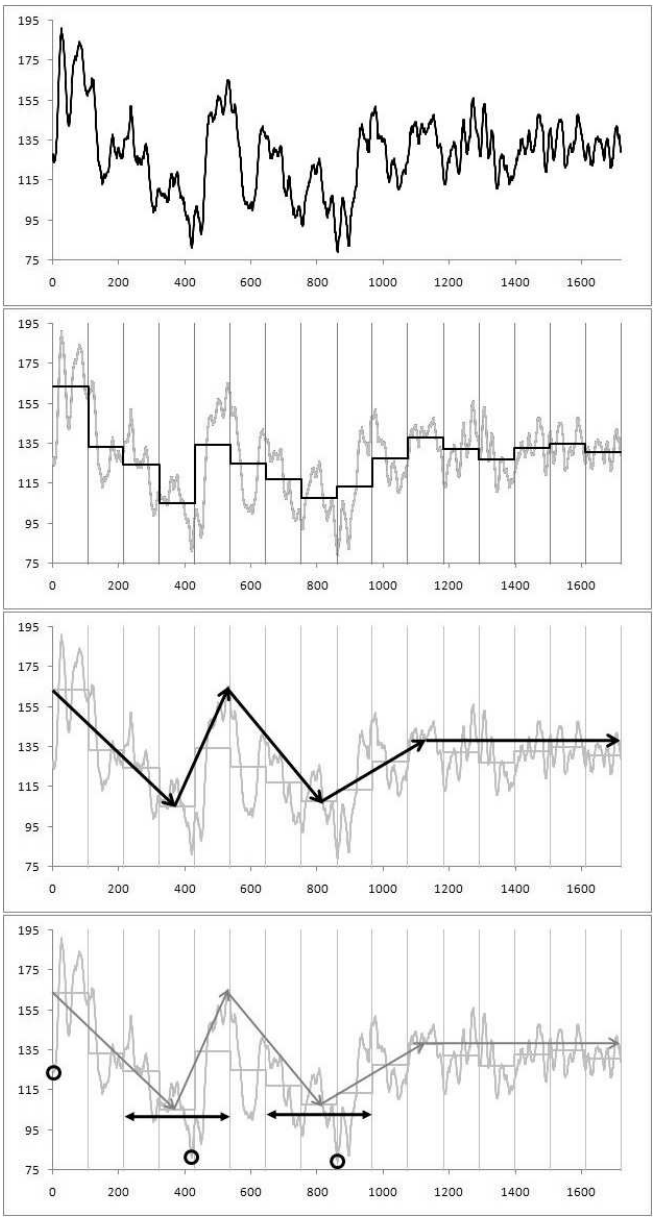,width=246pt}}
	\caption{\label{fig:beat} Implemented ``beat finder'' algorithm: 
	 1) one cycle (temporal signal), 
	 2) mean values of signal within the 16 ranges, 
 	 3) variation of averages, 
	 4) ranges (horizontal arrows) in which the minimum (encircled) represents a pulse.}
\end{figure}

Implemented algorithm, illustrated in Fig.~\ref{fig:beat}, is based on the following ideas:
The studied signal (one cycle) is divided into $n$ equal ranges.
The variation direction of mean values other each range is studied to detect when it changes 
from decreasing or constant to increasing (which corresponds to the notion of beginning of note);
The exact inflexion point is determined as the minimum in the temporal signal corresponding
to the inflexion range and on the two adjacent ranges.

Knowing these instants when acoustic events occur (called pulses), it is obvious to
find the minimal pulsation, so that each pulse and the total duration of a cycle are a multiple 
of this pulsation, and hence to reach the rhythmic suavity.
With such a definition, the rhythmic suavity is a real number lying between 0 and 1.

\medskip
\section{Stage 3: Tonal suavity}

We call ``chord'' the frequency content of the part of the signal
contained between two successive pulses.
A chord is constituted by a certain number of ``notes'' corresponding to frequencies
present in the chord.
	
A method derived from Euler's work will be used to compute the "beauty" of each
chord relatively to the tonal centre of the signal.

The tonal suavity for the whole acoustic signal will be obtained by balancing the beauty
of each chord with respect to its duration.

\subsection{Determination of the tonal centre}

When considering music, it is easy to determine the tonal centre (i.e. the tonic) with
the score.

But, when considering a noise, this task is not so simple.
Nevertheless, there are many ways to determine an ``equivalent note'' or a tonal centre of
an acoustic signal. 

It is possible to use prominent frequency methods: 
both  ``Tone to noise ratio'' as defined by standards\cite{bib-tnr,bib-pr}, 
and  ``Prominence ratio''defined by standard\cite{bib-pr} can be used.
For these methods, a tone is said to be prominent if it fulfills some criteria within
the critical band centered at the frequency of the tone.
When multiples tones are in the same critical band, prominence ratio is more effective; 
and when multiple tones exist in adjacent critical bands (strong harmonics), tone-to-noise 
ratio should be preferred.
From these frequencies, the most emerging one can be chosen as the tonal
centre.

In this work, we define the tonal centre as the fundamental frequency of the signal,
since it is generally a value of interest when considering a periodic signal (at least
from a mechanical and energetic point of view).
It means that we retain the lowest frequency peak in the signal.
For this purpose, a Bartlett function is used in the FFT in order to favor lower frequencies.

In the following, $f_0$ will denote the frequency corresponding to the
tonal centre of the signal.

\subsection{Tonal suavity}

Tonal suavity will be calculated using a modified version of Euler's works\cite{bib-euler1,bib-euler2,bib-euler3}.
This approach\cite{bib-euler4} is primarily a physicist's approach, and is close to Helmholtz's work.
Besides, the classification of beauty of chords constituted by only two notes are exactly the same 
for Euler\cite{bib-euler1} and Helmholtz\cite{bib-helm}. 

\medskip
This modified version of Euler's work has a major advantage: 
once the classification scale of the beauties of chords constituted by two notes is determined,
then it can be extended to chords constituted by any number of notes spread over any number of octaves.

A chord (previously defined as the frequency content between two successive pulses)
is constituted by $n$ notes.
We shall denote $f_1$, $f_2$, ..., $f_n$ the $n$ frequencies corresponding
of these $n$ notes of the chord ($f_0$ denoting the frequency of the
tonal centre of the signal as defined previously).

\subsection{Euler's work}

Euler defines a ``degree of sweetness'', a kind of ``easiness'', which
is his indicator: the less this indicator, easier to perceive the order between notes, 
or equivalently more beautiful the sound.

\medskip
Euler's theory can be summarized as follows:
\begin{itemize}
   \item A note is replaced by a number lying between 0 (unisson) and 12 (octave) and
	corresponding to its interval to the tonal centre.
   \item The ``exponent'' of a chord is defined as the LCM (least common multiple) of the ratios of frequencies
	related to the tonal centre of the signal:
	exponent = LCM($f_1/f_0, ..., f_n/f_0$).
	In order to calculate LCM, we have to manage only integers. Hence, if ratios
	$f_i/f_0$ are not all integers, they are all multipled by an adequate coefficient.
   \item The exponent is then decomposed in product of primes:
	\begin{equation}
	\label{eq-exp}
	   \text{exponent}_{\text{chord}} = \prod_{k} p_k^{j_k}
	\end{equation}
	where $k$ is the number of primes $p_k$ necessary to perform this decomposition and
	$j_k$ their exponent.
	This prime decomposition is given in table~\ref{tab-euler}.
	Some examples are proposed in table~\ref{tab-notes}.
   \item The easiness of a chord is computed from the previous prime decomposition as:
	\begin{equation}
	\label{eq-easy}
	   \text{easiness}_{\text{chord}} = 1+\sum_{k} j_k(p_k-1)
	\end{equation}
	The result is written using roman numbers in table~\ref{tab-notes}.
\end{itemize}

To illustrate the methods in this article, we use music, and more precisely, we use C Major (no key signature).
Hence a chord whose bass is C will be represented by the number 1 (because it is also the
tonal centre). A note C one octave higher will be represented by the number 2 (because its frequency
is the double of the one of tonal centre). A note F a fourth higher than the tonal centre will be
represented by 4/3 which represents the ratio of its frequency to the tonal centre.
These ratios between notes heights are given in table~\ref{tab-euler}.
\begin{table}
\caption{\label{tab-euler}Euler's music scale and easiness}
   \begin{ruledtabular}
      \begin{tabular}{cccc}
	interval                                      & $i$ & Prime decomposition     & easiness \\
	\hline
	unison                                        & 0    & 1                                   & 1 \\
	minor $\text{2}^{\text{nd}}$  & 1    &$2^4.3^{-1}.5^{-1}$  &11\\
	Major $\text{2}^{\text{nd}}$  & 2    &$2^{-3}.3^2$             & 8\\
	minor $\text{3}^{\text{rd}}$   & 3    &$2.3.5^{-1}$                & 8\\
	Major $\text{3}^{\text{rd}}$   & 4    &$2^{-2}.5$                   & 7\\
	$\text{4}^{\text{th}}$             & 5    &$2^2.3^{-1}$               & 5\\
	aug. $\text{4}^{\text{th}}$     & 6    &$2^{-5}.3^2.5$           &14\\
	$\text{5}^{\text{th}}$             & 7    &$2^{-1}.3$                   & 4\\
	minor $\text{6}^{\text{th}}$   & 8    &$2^3.5^{-1}$               & 8\\
	Major $\text{6}^{\text{th}}$   & 9    &$3^{-1}.5$                   &7\\
	minor $\text{7}^{\text{th}}$    &10   &$2^4.3^{-2}$              & 9\\	
	Major $\text{7}^{\text{th}}$    &11   &$2^{-3}.3.5$               &10\\
	octave                                         &12   &$2$                               & 2\\
      \end{tabular}

   \end{ruledtabular}
\end{table}
Some examples of numbering of intervals are given in table~\ref{tab-notes}.
\begin{table}
\caption{\label{tab-notes}Numbering of some chords according to Euler}
   \begin{ruledtabular}
      \begin{tabular}{lp{18mm}p{18mm}p{18mm}}
	\raisebox{6ex}{chord} &
	\makebox[15mm]{
	\begin{music}
	   \startextract
	         \notes \zqu{cj} \sk \enotes
      	   \endextract
	\end{music}
	}
	&
	\makebox[15mm]{
	\begin{music}
	   \startextract
	         \notes \zqu{cf} \sk \enotes
      	   \endextract
	\end{music}
	}
	&
	\makebox[15mm]{
	\begin{music}
	   \startextract
	         \notes \zqu{ceg} \sk \enotes
      	   \endextract
	\end{music}
	}
	\\
	ratio & 2 & 4/3 & 5/4 and 3/2 \\
	coefficient & 1 & 3 & 4\\
	\raisebox{6ex}{tonal centre} &
	\makebox[15mm]{
	\begin{music}
	   \startextract
	         \notes \zqu{c} \sk \enotes
      	   \endextract
	\end{music}
	}
	&
	\makebox[15mm]{
	\begin{music}
	\setclef{1}{60}
	   \startextract
	         \notes \zqu{F} \sk \enotes
      	   \endextract
	\end{music}
	}
	&
	\makebox[15mm]{
	\begin{music}
	\setclef{1}{60}
	   \startextract
	         \notes \zqu{C} \sk \enotes
      	   \endextract
	\end{music}
	}
	\\
	& unchanged &modified & modified\\
	exponent & LCM(1;2)=2 & LCM(3;4)=12 & LCM(4;5;6)\\
	                & $=2^1$ & $=2^2.3$ & $=2^2.3.5$\\
	easiness & 1+1 = II & 1+2+2 =IV & 1+2+2+4 =IX \\
      \end{tabular}
   \end{ruledtabular}
\end{table}

\subsection{Modification, extension}

Concerning Euler's work, we can make the following remarks:
\begin{itemize}
   \item In order to have only natural numbers, it is necessary to multiply the approximations of
	$f_i/f_0$ ratios according to table~\ref{tab-euler} by a coefficient.
	Doing this corresponds to change the tonal centre used to perform the numbering.
	For example, in second chord of table~\ref{tab-notes}, we have to multiply by 3 
	to have integers. This corresponds to change the C tonal centre by a F tonal centre two octaves lower.
   \item Nevertheless, insofar as the same coefficient is used to number all the chords of the signal,
	this has only a relative importance because the numbering is still done with respect
	to the same note even if it is no more the tonal centre.
\end{itemize}

\medskip
Because the fundamental frequency is a value on interest in the industrial
cases we face, we do not want to change the tonal centre $f_0$ used in the numbering 
process. It is therefore necessary to modify the approximations of intervals.
We have to find a set of primes so that i) it is possible to correctly approximate
the powers of twelfth root of 2, ii) having only powers of 2 as denominators and iii) so that
easiness obtained for the intervals are classified in the same order of the ones obtained by Euler.

Such an approximation is presented in table~\ref{tab-VMt}.
\begin{table}
\caption{\label{tab-VMt}Proposed scale and easiness}
   \begin{ruledtabular}
      \begin{tabular}{cccc}
	interval                                       & $i$ & Prime decomposition    & easiness \\
	\hline
	unison                                        & 0    &$1$                                & 1 \\
	minor $\text{2}^{\text{nd}}$  & 1     &$2^{-12}.3.5.7.41$    &65\\
	Major $\text{2}^{\text{nd}}$  & 2    &$2^{-7}.11.13$            &30\\
	minor $\text{3}^{\text{rd}}$   & 3    &$2^{-4}.19$                 &23\\
	Major $\text{3}^{\text{rd}}$   & 4    &$2^{-6}.3.5$               &17\\
	$\text{4}^{\text{th}}$             & 5    &$2^{-8}.7^3$              &27\\
	aug. $\text{4}^{\text{th}}$     & 6    &$2^{-10}.31.47$         &87\\
	$\text{5}^{\text{th}}$             & 7    &$2^{-5}.3^2.5$           &14\\
	minor $\text{6}^{\text{th}}$   & 8    &$2^{-8}.11.37$            &55\\
	Major $\text{6}^{\text{th}}$   & 9    &$2^{-9}.3.17^2$         &44\\
	minor $\text{7}^{\text{th}}$   &10   &$2^{-12}.3^2.5.7.23$ &49\\
	Major $\text{7}^{\text{th}}$  &11    &$2^{-9}.3.11.29$         &50\\
	octave                                        &12   &$2$                                &2\\
      \end{tabular}
   \end{ruledtabular}
\end{table}
Contrary to Euler, who only used powers of  2, 3 and 5 
for the decompositions, we use 2, 3, 5, 7, 11, 13, 17, 19, 23, 29, 31, 37, 41 and 47.
Nevertheless, it does not make the computation algorithm more complicated (since prime decomposition
is not computed but directly obtained by construction).

Easiness of intervals corresponding to this new decomposition have higher values than
Euler's ones, but globally respect the same order.
Some differences can be seen which permit to include Euler's latest work\cite{bib-euler2}, and 
to better take into account some consonances.
Spreading values of easiness over a large range yields more distinction in suavity marks.

\subsection{Implemented algorithm}

The algorithm is as follows:
\begin{enumerate}
   \item Determination of the notes constituting a chord:

	From the spectrum of the frequency content corresponding to a chord (signal contained between
	two acoustic events), we take the $n$ lowest frequency peaks.
	In developed program, $n\le10$, and we stop looking for peaks at 
	$\min(8000~\text{Hz}, \text{Nyquist frequency})$.
	We then have $n$ frequencies $f_n$ describing the chord.

   \item Approximation of the notes of the chord:

	Each frequency $f_n$ si approximated by $f_n\approx f_0.2^{k_n}.(\sqrt[12]{2})^{i_n}$,
	where $k_n$ is the number of octaves between $f_n$ and $f_0$ and $i_n$  
	the halftone which is the closest to $f_n$ in the considered octave.
	$(\sqrt[12]{2})^{i_n}$ is then replaced by its approximation according to 
	tables~\ref{tab-euler} or \ref{tab-VMt}.

   \item  Calculation of the chord's exponent using equation (\ref{eq-exp}).

   \item Calculation of the chord easiness using equation (\ref{eq-easy}).

   \item Normalization:

	The lower the easiness, the more the chord is tuneful.
	On the contrary, we want that the higher the suavity, the more the chord is tuneful.
	We also want the suavity to lie between 0 and 1.
	
	We can notice that, for a given chord of $n$ notes:
	$
	1 \le \text{easiness} \le b
	$
	with:
	\begin{equation}
	\label{eq-bn}
	   b= o_n + a_n
	\end{equation}
	where $o_n$ is the upper bound of the number of octaves between the tonal centre and the notes, 
	and $a_n$ is the upper bound of the easiness of a chord constituted by $n$ notes.

	$a_n$ is the upper bound of equation (\ref{eq-easy}), which is calculated by taken the maximum of each 
	term of the sum:
	\begin{equation}
	\label{eq-an1}
	\begin{array}{rl}
	   a_n =& 1 + \sum_k \left(\max_{i=1}^{12} j_i-\min_{i=1}^{12} (0,j_i)\right)(p_k-1)\\[+1ex]
	         =&28 \text{~if using table~\ref{tab-euler}}\\
	         =&318 \text{~if using table~\ref{tab-VMt}}
	\end{array}
	\end{equation}
 
	If we only consider frequencies between 1 and 10~kHz, then we have a maximum of
	14 octaves, 
	which means that $o_n=14$ can be used as upper bound.

   \item Calculation of the chord suavity:

	Finally, the suavity is defined by:
	\begin{equation}
	\label{eq-suachord}
	   \text{suavity}_{\text{chord}} = 1 - \dfrac{\text{easiness}-1}{b-1}
	\end{equation}
	which lies between 0 and 1 (1 being the better).
   \item We finally compute the tonal suavity of the signal as the balanced average of the
	chords suavity with the duration of the chords:
	\begin{equation}
	\label{eq-suaton}
	   \text{suavity}_\text{tonal} = \sum_i  \dfrac{
               \text{suavity}_{\text{chord}_i} .
   	   \text{duration}_{\text{chord}_i}
	   }{\text{total duration of the signal}}
	\end{equation}
	From its definition, the tonal suavity lies between 0 and 1.
\end{enumerate}

\section{Stage 4: Harmonic suavity}

Harmonic suavity represents the beauty of the transition between two successive chords 
constituting the signal (plus the last chord followed by the first one for periodic signals).
More precisely, harmonic suavity quantifies the beauty of the part of the second chord
which does not come from a transformation (belonging to the dihedral group) of the
first chord.
In this sense, harmonic suavity is related to the rate of change between two acoustic events.

\subsection{\label{sec:existant}Existing works}

In his work, Euler extends his method in order to determine the beauty of a set of
chords or to a whole musical work.

The main idea is the following: the transition between two chords is
harmonious if the chord composed of all notes composing the two chords is
itself harmonious.
By extension, this method could be applied to any number of chords, and hence
to a whole musical work.

\begin{figure}
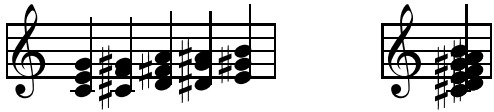

\parbox{246pt}{%
   ~\hspace{-2cm}~
   \begin{music} 
      \startextract
         \notes \zqu{ceg} \sk \zqu{^cf^g} \sk \zqu{d^fh} \sk \zqu{^dg^h} \sk \zqu{e^gi} \sk\enotes
      \endextract
   \end{music}%
   ~\hspace{-5cm}~
   \begin{music} 
      \startextract
         \notes \zqu{c^ceg^gi}\rq{d^df^fh^h} \sk \enotes
      \endextract
   \end{music}%
}
\caption{\label{fig:marche}An ascending sequence}
\end{figure}

Two mains objections can be done:
\begin{enumerate}
   \item Such a method yields a bad mark for transposed chords, as in sequences.
	Fig~\ref{fig:marche} exhibits a perfect chord C--E--G, i.e. the C major triad in root position,
	as the starting chord of a ascending sequence by successive halftones steps. 
	Such a sequence of chords is absolutely not chocking to the ears, because
	a motif is immediately recognized, which is repeated and transposed.
	With Euler's method, we have to take into consideration the chord
	C--C\#--D--D\#--E--F--F\#--G--G\#--A--A\#--B, i.e. constituted by
	all the halftones: this chord has a high easiness or a bad tonal suavity, which
	perfectly agree with what we hear.
   \item Euler does not calculate the beauty of the transition between two chords, but 
the overall beauty of two (or more) chords considered as a single entity.
\end{enumerate}

\medskip
Another possible way consists to use musicology rules related to the beauty of
transition of chords (which is sometimes referred to as ``chords in movement'').
In this approach, chords are qualified (stables or attractive) and the movement
of the bass is analyzed with respect to the degree of the chord (i.e. the position of
the bass relatively to the tonal centre).
Rules are numerous and delicates. They need an analysis, which can be qualified as
syntactic. It is quite impossible to program such rules, especially if we want to
number chords having more than 6 notes (which cannot be taken into account)...

\subsection{Developed approach}

To expose our approach, we continue to explain it using musical example, even
if our goal remains the study of industrial noises.

We propose to simplify Euler's work and combine it with some rule
of musical harmony.
From musical harmony we keep numbering a chord relatively to its bass, without
taking into account octaves.
From Euler's work, we keep the idea according to which the transition between two 
chords is beautiful if the chord made of all notes of both chords is beautiful.
But the method is only applied to the transition between two chords.
We only consider a chord relatively to the bass of the previous one (and no more relatively
to the tonal centre). 
Finally, we solve the problem described previously concerning sequences by ``suppressing''
from the second chord the part that ``comes from the first one'' (in a sense of
transformations belonging to the dihedral group).

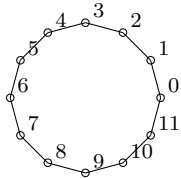
\begin{figure}
   \begin{center}
      \begin{picture}(2,2)(-1,-1)%
          \VMlinedodec
          \VMcircdodec
         \VMgraddodec
      \end{picture}%
   \end{center}
\caption{\label{fig:dodec0} Dodecagone of the 12 halftones}
\end{figure}

\medskip
As aforementioned, considering a note independently from its octave and relatively to a reference 
note is equivalent to consider its interval to the reference note.
Hence a note can be represented by an elements of the cyclic group $\Z/12\Z$.
This cyclic group can be viewed as a regular polygon with $12$ sides,
i.e. a dodecagon: the 12 apexes represent the 12 possible halftones for a 
note (relatively to the bass of the chord). It is depicted in figure Fig.~\ref{fig:dodec0}.

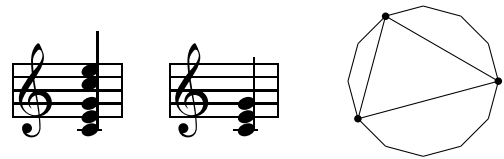
\begin{figure}
\parbox{246pt}{%
   ~\hspace{-4cm}~
   \begin{music} 
      \startextract
         \notes \zqu{cegjl} \sk \enotes
      \endextract
   \end{music}%
   ~\hspace{-7cm}~
 \begin{music} 
      \startextract
         \notes \zqu{ceg} \sk \enotes
      \endextract
   \end{music}%
   ~\hspace{-3cm}~
   \begin{picture}(2,2)(-1,-1)%
       \VMlinedodec
      \VMnote0
      \VMnote4
      \VMnote7
      \VMLine04
      \VMLine47
      \VMLine70
   \end{picture}%
}

\caption{\label{fig:dodec1}Representations of a chord (Considered chord / compact chord / corresponding polygon)}
\end{figure}

\medskip
In the following, notes will be considered independently  from octaves:
hence C--E--G--C--E is equivalent to C--E--G, i.e. the chord is presented in a ``compact'' 
form, as illustrated in  figure Fig~\ref{fig:dodec1}.

Considering a chord made of $p$ notes, it can be represented by a $p$ sided 
polygon whose apexes belong to the dodecagon.
A set $T$ of  $p$ values in $\Z/n\Z$, which is equivalent to a $p$ sided polygon
whose apexes belong to an $n$ sided polygon, is hence equivalent to a
chord constituted by $p$ notes.

For example $T=(0,4,7)$ represents in $\Z/12\Z$ the chord and the polygon depicted in figure
Fig~\ref{fig:dodec1}.
Chord $T=(0,4,7)$ correspond to any chord whose second note is located 4 halftones 
from the first one, and whose third note is located 7 halftones from the first one, the first
one being its bass.
Such a chord can be C--E--G, as in figure Fig.~\ref{fig:dodec1}, but also any chord being a
transposition of it (see figure Fig.~\ref{fig:marche}).

\medskip
The dihedral group $D_n$ is a group of order $2n$ of plane isometries letting the regular $n$ sided
polygon unchanged.
$D_n$ is made of $n$ rotations and $n$ symmetries.
It is generated by two elements: the rotation of one vertice to the next $R_1$, and the symmetry 
with respect to abscissa axe $S_0$.
The dihedral group $D_n$ can hence be written in the form $D_n=\{R_0=Id, R_1, ..., R_1^{n-1}, S_0,
R_1\circ S_0, ..., R_1^{n-1}\circ S_0\}$.
From a computation point of view, it means that it is sufficient to dispose of $R_1$ and $S_0$ 
transformations to be able to browse the entire dihedral group.

\medskip
From a musical (and a frequential) point a view, we can notice that:
\begin{itemize}
   \item $R_0=Id$ corresponds to a transposition of chord $T$ (as in a sequence);
   \item $R_i$, with $i\ge 1$, corresponds to an inversion of chord $T$ (same notes in a different order)
	 with eventually a transposition;
   \item $S_0$ corresponds to a retrograde of chord $T$ (chord made of the same intervals in the opposite
	direction)
	 with eventually a transposition;
   \item $R_i\circ S_0$, with $i\ge 1$, corresponds to the retrograde of an inversion of chord $T$
	 with eventually a transposition;
\end{itemize}

\medskip
Although each transformation of the dihedral group is different from all others, it does
not imply that it exists an unique decomposition of the transformation of a polygon $T$ 
into a polygon $V$ when $p<n$.

Consider for example chord $T=(0,5,10)$ corresponding to chord C--F--A\# and
chord $V=(0,2,7)$ corresponding to chord D--E--A.
Then $V=R_2(T)$, but also $V=S_0(T)$.

When several transformations exist, then we choose the one which comes first
in the order of transformations, when classified as previously:
$D_n=\{R_0=Id, R_1, ..., R_1^{n-1}, S_0,
R_1\circ S_0, ..., R_1^{n-1}\circ S_0\}$.
It just means that it is easier to hear a transposition, then an inversion, 
then a retrograde chord, and finally the retrograde chord of an inversion.

\medskip
Finally, if such a transformation exists, then it is possible to give a ``beauty mark'' to 
the transformation.
We chose the following marks: 1 for $R_0=Id$, 0.9 for $R_i$ ($i\ge1$), 0.8 for $S_0$ 
and 0.7 for $R_i\circ S_0$ ($i\ge1$).

\subsection{Transformation from a chord to another -- dihedral component 
$\VMdihedral$ and non congruous component $\VMreste$}

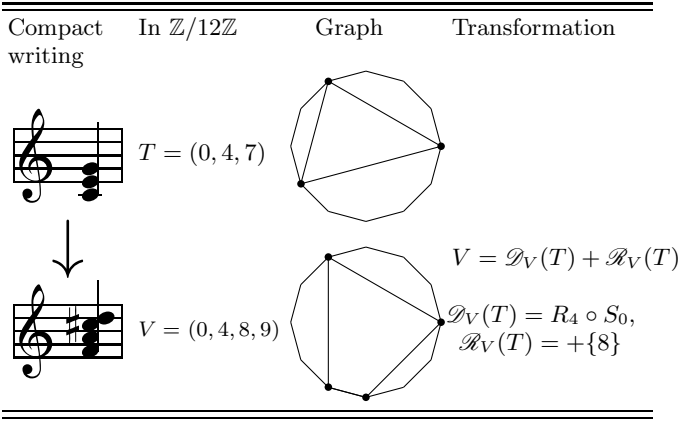
\begin{figure}
\parbox{246pt}{%
   \begin{ruledtabular}
      \begin{tabular}{p{16mm}lp{2cm}l}
	Compact writing & In $\Z/12\Z$ &~~~Graph & Transformation
	\\
	\makebox[15mm]{
	\begin{music}
	   \startextract
	         \notes \zqu{ceg} \sk \enotes
      	   \endextract
	\end{music}
	}
	&
	\raisebox{6ex}{$T=(0,4,7)$}
	&
	\begin{picture}(2,2)(-1,-1)%
       	   \VMlinedodec
	   \VMnote0
      	   \VMnote4
	   \VMnote7
	   \VMLine04
	   \VMLine47
	   \VMLine70
   	\end{picture}
	&
	\\[-1ex]
	{\centerline{\Huge$\downarrow$}}&&&$V=\VMdihedral_V(T)+\VMreste_V(T)$
	\\[-6ex]
	\makebox[15mm]{
	\begin{music}
	   \startextract
	         \notes \zqu{fh^j}\rq{k} \sk \enotes
      	   \endextract
	\end{music}
	}
	&
	\raisebox{6ex}{\footnotesize$V=$ $(0,4,8,9)$}
	&
	\begin{picture}(2,2)(-1,-1)%
       	   \VMlinedodec
	   \VMnote0
      	   \VMnote4
	   \VMnote8
	   \VMnote9
	   \VMLine04
	   \VMLine48
	   \VMLine89
	   \VMLine90
   	\end{picture}
	&
	\hspace{-1em}
	\raisebox{6ex}{\parbox{28mm}{$\VMdihedral_V(T)=R_4\circ S_0$,\\ 
	$\VMreste_V(T)=+\{8\}$}}
	\\
   \end{tabular}
\end{ruledtabular}
}
\caption{\label{fig:VT} Comparison between chords}
\end{figure}

Consider chord $T=(0,4,7)$, which may represent the perfect chord
C--E--G already presented in figure Fig.~\ref{fig:dodec1}.
Let $V=(0,4,8,9)$ be the chord F--A--C\#--D, as illustrated in figure Fig.~\ref{fig:VT}.

It is easy to see that $R_4(S_0(T))=(0,4,9)=V-\{8\}$.

We call dihedral component, the element of $D_n$ allowing to transform $T$ into $V$.
It is denoted $\VMdihedral_V(T)$.
In this case, the dihedral component is equal to $\VMdihedral_V(T)=R_4\circ S_0$.

We also define the non congruous component of $V$ relatively to $T$, and
we denote it $\VMreste_V$, as the remainder when ``subtracting'' from $V$ 
its dihedral component $\VMdihedral_V(T)$: it corresponds to the notes belonging
to $V$ which are not  issued from the transformation of $T$.
This definition is quite natural, and we reach: $\VMreste_V(T)=+\{8\}$.

\medskip
When no transformation exists between two chords, then the dihedral component
$\VMdihedral=\emptyset$ and the non congruous component $\VMreste$ is equal 
to the second chord.

\subsection{Arithmetic of transformations}

Consider a chord $T$ and its successor $V$.
As explained, a way to write that $V$ issued from a transformation of $T$ is:
\begin{equation}
\label{eq-decomp}
V=\VMdihedral_V(T) + \VMreste_V(T)
\end{equation}

Equation (\ref{eq-decomp}) can be seen as a decomposition or a division of $V$ by $T$: 
$V$ represents the dividend, $T$ the divisor, $\VMdihedral_V(T)$ the quotient
and $\VMreste_V(T)$ the remainder.
This equation could also be written in a modular arithmetic form as:
$V \equiv \VMreste_V(T) \mod [\VMdihedral_V(T)]$.

\medskip
Equation (\ref{eq-decomp}) can be written in inverse form:
\begin{equation}
\label{eq-decompinv}
T=\VMdihedral_T(V) + \VMreste_T(V)
\end{equation}
with:
\begin{equation}
\label{eq-inv}
\left\{
\begin{array}{rcl}
   \VMdihedral_T &=& \VMdihedral_V^{-1}\\
   \VMreste_T &=& -\VMdihedral_V^{-1}.\VMreste_V = -\VMdihedral_T(\VMreste_V)
\end{array}
\right.
\end{equation}

Since any representative of the dihedral group can only be written as $R_i$ or
$R_{i}\circ S_0$, then (indexes belong to $\Z/n\Z$):
\begin{equation}
\label{eq-invdied}
\left\{
\begin{array}{lrcl}
   \text{if } \VMdihedral_V=R_i, &\VMdihedral_T &=& R_{-i}\\[+1ex]
   \text{if } \VMdihedral_V=R_{i}\circ S_0, &\VMdihedral_T &=& S_0 \circ R_{-i}\\
\end{array}
\right.
\end{equation}
From equation (\ref{eq-invdied}), we can notice that the ``beauty mark''
of the transformation is the same for the direct and the reverse transformation.

\begin{figure}
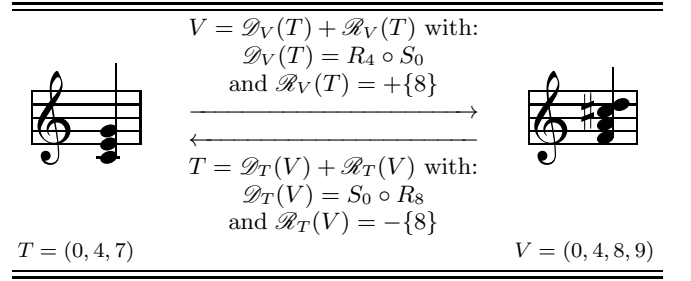

   \begin{ruledtabular}
      \begin{tabular}{p{18mm}cp{19mm}}
	\raisebox{-5ex}{
	\makebox[15mm]{
	\begin{music}
	   \startextract
	         \notes \zqu{ceg} \sk \enotes
      	   \endextract
	\end{music}
	}
	}
	&
	\begin{tabular}{c}
	 $V=\VMdihedral_V(T)+\VMreste_V(T)$ with:\\
	$\VMdihedral_V(T)=R_4\circ S_0$\\ and $\VMreste_V(T)=+\{8\}$
	\\
	\rightarrowfill
	\\
	\leftarrowfill
	\\
	$T=\VMdihedral_T(V)+\VMreste_T(V)$ with:\\
	$\VMdihedral_T(V)=S_0\circ R_8$\\ and $\VMreste_T(V)=-\{8\}$
	\end{tabular}
	&
	\raisebox{-5ex}{
	\makebox[15mm]{
	\begin{music}
	   \startextract
	         \notes \zqu{fh^j}\rq{k} \sk \enotes
      	   \endextract
	\end{music}
	}}
	\\
	{\footnotesize $T=(0,4,7)$}
	&&
	{\footnotesize	$V=(0,4,8,9)$}
	\\
   \end{tabular}
\end{ruledtabular}
\caption{\label{fig:TVT} Arithmetic between chords}
\end{figure}

\medskip
We study again transformation from $T$ into $V$ and from $V$ into $T$ 
defined by $T=(0,4,7)$ C--E--G and $V=(0,4,8,9)$ F--A--C\#--D and illustrated
in figure Fig~\ref{fig:VT}.
Dihedral and non congruous components can be calculated directly as in previous 
section, or using equations (\ref{eq-inv}) and (\ref{eq-invdied}).
Results are reported in figure Fig~\ref{fig:TVT}.

In this case we have:
\begin{itemize}
   \item Direct calculation from $T$ to $V$ gives  
	$\VMdihedral_V(T)=R_4\circ S_0$ and $\VMreste_V(T)=\{8\}$,
   \item Direct calculation from $V$ to $T$ gives  
	$\VMdihedral_T(V)=R_4\circ S_0$ and $\VMreste_T(V)=-\{8\}$.
   \item Equations (\ref{eq-inv}) and (\ref{eq-invdied}) give
	$\VMdihedral_T(V)=S_0\circ R_{-4}=S_0\circ R_8$
	and $\VMreste_T(V)=-(S_0\circ R_8)(\{8\})=-S_0(\{4\})=-\{8\}$.
\end{itemize}
The writing of $\VMdihedral_T(V)$ is not unique, but 
$[R_4\circ S_0](T) = [S_0\circ R_8](T)$.
The most important point remains that the beauty mark of the dihedral
component remains the same, and that both decompositions lead to
the same non congruous component $\VMreste_T(V)$.

\subsection{Harmonic suavity of the transition between two chords}

The harmonic suavity, corresponding to the beauty of the transition from chord $T$
to chord $V$, will be calculated as follows:
\begin{itemize}
   \item If it exists a transformation from $T$ to $V$ then:
   \begin{itemize}
	\item if the number of notes of $V$ is less or equal to the number of notes
	of $T$ (i.e. the non congruous component is less or equal to zero), then
	the harmonic suavity is equal to the mark of the transformation: 
	1 for $R_0$, 0.9 for $R_i$ ($i\ge1$), 0.8 for $S_0$ and 0.7 
	for$R_i\circ S_0$ ($i\ge1$);
	\item if the number of notes of $V$ is strictly greater than the number of notes
	of $T$ (i.e. the non congruous component is strictly greater than zero), then
	we calculate the tonal suavity of chord $\VMreste_V(T)$ with respect to the
	bass of $T$ (i.e. we replace $f_0$ by $f_1(T)$ in stage 3 algorithm);
   \end{itemize}
   \item If there is no transformation from $T$ to $V$ then:
	we calculate the tonal suavity of chord $V$ (we recall that in this case $\VMreste_V(T)=V$) 
	with respect to the bass of $T$ (i.e. we replace $f_0$ by $f_1(T)$ in stage 3 algorithm).
\end{itemize}
In cases where stage 3 (tonal suavity) is involved (i.e. when the non congrusous component si strictly
greater than zero), the normalization of the tonal suavity is done using 
equation (\ref{eq:bn}) with $o_n=0$ (because using a compact notation, we only 
consider one octave), and $a_n$ is unchanged.

\medskip
La figure~\ref{fig:fin} reprend l'exemple de la figure~\ref{fig:TVT} et donne les accords r\'esultant
de la transition dans le cas o\`u la m\'ethode d'Euler aurait \'et\'e conserv\'ee et dans la m\'ethode propos\'e.

Figure Fig.~\ref{fig:fin} use the same example as in figure~\ref{fig:TVT} and gives
the resulting chords to the studied transition obtained using Euler's work as explained in 
section  \ref{sec:existant}  and using the exposed method for comparison.
\begin{figure}
   \begin{ruledtabular}
      \begin{tabular}{p{38mm}p{38mm}}
	\multicolumn{2}{c}{
	\makebox[25mm]{
	\begin{music}
	   \startextract
	         \notes \zqu{ceg} \sk \sk \zqu{fh^j}\rq{k} \sk \enotes
      	   \endextract
	\end{music}
	}
	}\\
	Euler's method: & Proposed method:\\
	tonal suavity of the sum of the chords is calculated. 
	No difference is made between beauty of the chords and of their transition.
	Calculation of the tonal suavity of chord:
	&
	Tonal suavity of each chord has been calculated in previous stage.
	Harmonic suavity of the transition is calculated as the tonal suavity 
	of the resulting chord:
	\\
	\multicolumn{1}{c}{
	\raisebox{-9ex}{
	\makebox[15mm]{
	\begin{music}
	   \startextract
	         \notes \zqu{=ceg^j}\rq{fhk} \sk \enotes
      	   \endextract
	\end{music}
	}}}
	&
	\multicolumn{1}{c}{
	\raisebox{-9ex}{
	\makebox[15mm]{
	\begin{music}
	   \startextract
	         \notes \zqu{c_h} \sk \enotes
      	   \endextract
	\end{music}
	}}}
	\\
   \end{tabular}
\end{ruledtabular}
\caption{\label{fig:fin} Resulting chords}
\end{figure}

\subsection{harmonic suavity of a signal}

The harmonic suavity of a signal is defined as the average of harmonic
suavity of all transitions of chords constituting the signal.

For periodic signal, we also have to take into account the transition between
the last chord and the first chord of the signal.

\section{Stage 5: Global suavity}

The global suavity synthesizes the three previous indicators to give a global mark of the
beauty of the studied acoustic signal.

\begin{figure}
\begin{center}
\begin{picture}(4,2)(-2,-1)
   \Line(0,0)(0,1)
   \Line(0,0)(-0.8660,-0.5)
   \Line(0,0)(0.8660,-0.5)
   \put(0,0.5){\circle*{0.1}}%
   \put(-0.2598,-0.15){\circle*{0.1}}
   \put(0.3464,-0.2){\circle*{0.1}}
   \Line(0,0.5)(-0.2598,-0.15)
   \Line(-0.2598,-0.15)(0.3464,-0.2)
   \Line(0.3464,-0.2)(0,0.5)
   \put(-1,1.1){rhythmic: 50\%}
   \put(-2,-0.8){tonal: 30\%}
   \put(0.5,-0.8){harmonic: 40\%}
\end{picture}

Global suavity = 15.67\% (ratio of areas)
\end{center}
\caption{\label{fig:radar} Representation of global suavity}
\end{figure}
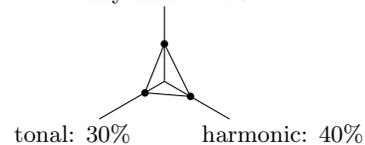

In order not to mix temporal information and frequency ones, a radar representation is used
(instead of the simple multiplication of suavity components). 
The ratio of the area of the triangle made by suavities divided by the maximum triangle area is 
used as  the global suavity, as illustrated in figure Fig.~\ref{fig:radar}.

\section{\label{sec:ex} Industrial examples}

In this section, we present some results obtained on industrial noises.

We use the first example in section \ref{sec:aix} to detail the calculation stages
in order to numerically illustrate presented algorithms.

Only results are presented concerning the second example.

\medskip
To perform analysis, a prototype software has been developed in fortran 77\cite{bib-g77},
with a graphical interface in japi\cite{bib-japi}.
This software has been developed only for research purpose, without any commercial
goal.

\subsection{\label{sec:aix} Asymmetrical two-stroke engine}

The signal is the noise emitted by an asymmetrical two-stroke engine.
One explosion happens at 1/4 of the first crankshaft rotation, while the second explosion
occurs at 3/4 of the first crankshaft rotation. No explosion happens during the second
rotation of the crankshaft.
The recorded signal is shown in figure Fig.~\ref{fig:typique}.

The software permits to identify the cycles.
It is found that the mean duration of cycles is 0.109~s with a standard deviation 
of 2.149.$10^{-3}$~s.
A ``cycle'', as detected by the program, correspond to a period of the signal,
which is, in this case, equal to 2 crankshaft rotations.
It correspond to an engine rotation speed equal to 1100~rpm.

\begin{figure}
   \centerline{\epsfig{figure=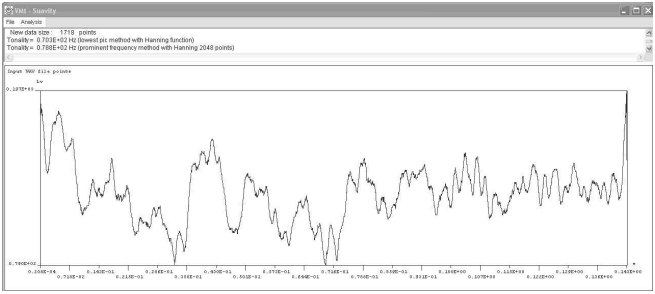,width=\columnwidth}}
   \caption{\label{fig:aixcycle}One cycle extracted from the signal (2 cranckshaft rotations)} 
\end{figure}

We only retain one single cycle (see figure Fig.~\ref{fig:aixcycle}) to continue the analysis.

\begin{figure}
   \centerline{\epsfig{figure=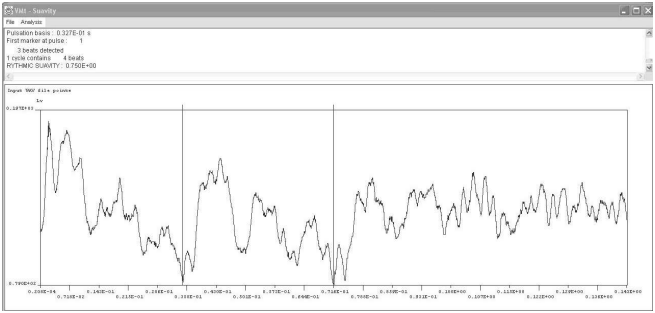,width=\columnwidth}}
   \caption{\label{fig:aixpulses}Detected pulses}
\end{figure}

Pulse analysis performed on one cycle (one period of the signal, 2 crankshaft rotations)
leads to pulses marked by vertical lines in figure Fig.~\ref{fig:aixpulses}.
It is found that 3 acoustic events occur at points 1, 420 and 862 during a cycle 
containing a total of 1718 points.
Algorithmic details are illustrated in figure~Fig.\ref{fig:beat}.
It corresponds to acoustic events occurring at pulses 0, 1 and 2 during a cycle containing 4 pulses.
The rhythmic suavity of the signal is equal to  $3/4 = 0.75$.

The periodicity of explosions suggests that the arrhythmia of the signal should be greater.
Due to only 2 explosions in one period (2 explosions separated by half a rotation of the crankshaft, 
then no explosion during 1.5 rotation), we thought that the rhythmic suavity would be
equal to $2/4=0.5$ (2 acoustic events for a total of 4 pulses).

But from figure Fig.~\ref{fig:aixcycle}, signal analysis clearly shows 3 areas:
1) a first area containing 3 decreasing peaks; 
2) an ascent followed by 3 decreasing peaks;
3) a last ascent followed by several peaks without special emergence.
Thus having found 3 acoustic events is correct.
In fact, mechanical noises are emitted at 1/4 of the second crankshaft rotation without
any explosion.
The level of this third pulse is anyway less than the one of the 2 previous pulses, but
its emergence is clearly present in the signal.

The analysis of durations between acoustic events agrees with the fact that a cycle
is made of 4 pulses.
The calculated rhythmic suavity (equal to $3/4$) is the correct one.

\medskip
Tonal centre is chosen as the fundamental frequency $f_0=70.3$~Hz.

\medskip
\begin{table}
   \caption{\label{tab-aixchord}Chord 1}
   \begin{ruledtabular}
      \begin{tabular}{lccll}
	frequencies (Hz)& $k_n$ & $i_n$ & Euler & Proposed \\
	&&&table~\ref{tab-euler}&table~\ref{tab-VMt}\\
	\hline
	$f_1=187.5$     & 1 & 5 & $2^33^{-1}$        & $2^{-7}7^3$\\
	$f_2=562.5$     & 3 & 0 & $2^3$                   & $2^{3}$\\
	$f_3=937.5$     & 3 & 9 & $2^33^{-1}5^1$ & $2^{-6}3^117^2$\\
	$f_4=1406.25$ & 4 & 4 & $2^25^1$            & $2^{-2}3^15^1$\\
	$f_5=1593.75$ & 4 & 6 & $2^{-1}3^25^1$ & $2^{-6}31^147^1$\\
	$f_6=1875.0$   & 4 & 9 & $2^43^{-1}5^1$ & $2^{-5}3^117^2$\\
	$f_7=2250.0$   & 5 & 0 & $2^5$                   & $2^5$\\
	$f_8=2718.75$ & 5 & 3 & $2^63^15^{-1}$ & $2^119^1$\\
	$f_9=3000.0$   & 5 & 5 & $2^73^{-1}$        & $2^{-3}7^3$\\
       $f_{10}=3468.75$ & 5 & 7 & $2^43^1$            & $3^25^1$\\
	~\\
	\multicolumn{3}{r}{easiness} & 25 & 169\\
	\multicolumn{3}{r}{coefficient} & 30 & 128\\
	\multicolumn{3}{r}{suavity} & 41.46\% & 49.24\%
      \end{tabular}
   \end{ruledtabular}
\end{table}
Table~\ref{tab-aixchord} shows the frequency content of chord 1 (first column), 
$k_n$ and $i_n$ factors corresponding to stage 3  algorithm, and the prime
decomposition according to Euler and proposed methods.
Doing the same computation for the second chord yields a coefficient in Euler's method equals to 120 to
obtain only integers. 
For the third chord, this coefficient in Euler's method remains equal to 120.
In order to calculate the tonal suavity of the whole cycle, we have to consider only one value of this 
coefficient (which correspond to a shift of the tonal centre), which is obviously the greater value (in order 
to have only integers). 
Updating this coefficient to 120 in the calculation of the first chord leads to a tonal
suavity of chord 1 equal to 36.58\%.
This updated value will be used in the computation of the averall tonal suavity.

Finally, tonal suavity (for the signal, i.e. for all chords) based on the proposed prime decomposition 
is equal to 36.02\% and to 34.10\% based on Euler's decomposition.

\medskip
The signal is composed of 3 chords. Frequencies of first chord have been presented in
table~\ref{tab-aixchord}. Frequency contents of chords 2 and 3 are given in
table~\ref{tab-aixchords}.
\begin{table}
   \caption{\label{tab-aixchords} Frequency content of chords 2 and 3 (Hz)}
   \begin{ruledtabular}
      \begin{tabular}{ll}
	Chord 2 & Chord 3\\
	\hline
	$f_1=281.25$    	&$f_1=140.625$\\
	$f_2=937.5$      	&$f_2=234.375$\\
	$f_3=1687.5$    	&$f_3=468.75$\\
	$f_4=2343.75$ 	&$f_4=562.5$\\
	$f_5=3093.75$  	&$f_5=890.625$\\
	$f_6=3656.25$  	&$f_6=1171.875$\\
	$f_7=4406.25$      	&$f_7=1546.875$\\
	$f_8=5062.5$        	&$f_8=1828.125$\\
	$f_9=5812.5$        	&$f_9=1921.825$\\
	$f_{10}=6468.75$	&$f_{10}=2203.125$\\
      \end{tabular}
   \end{ruledtabular}
\end{table}
\begin{table}
   \caption{\label{tab-aixchordss}Chords in compact form in $\Z/12\Z$}
   \begin{ruledtabular}
      \begin{tabular}{ccc}
	Chord1 & Chord 2 & Chord 3\\
	\hline
	0   &0 & 0\\
	1   &1 & 1\\
	3   &2 & 6\\
	4   &4 & 8\\
	7   &6 & 9\\
	10 &7 & \\
	11 &8 & \\
	    &9 & \\
      \end{tabular}
   \end{ruledtabular}
\end{table}
From chords written in $\Z/12\Z$ in compact form, as given in table~\ref{tab-aixchordss},
it appears that there is no transformation from chord 1 to chord 2, nor from chord 3 to
chord 1 (periodic signal), and chord 3 = Id(chord 2) -\{2, 4, 7\}.

In this case, analysis of chords transitions leads to:
\begin{itemize}
   \item chord 1 to chord 2: harmonic suavity = tonal suavity of chord 2 with respect to the bass of chord 1 
	(i.e. replacing $f_0$ by 187.5);
   \item chord 2 to chord 3: harmonic suavity = 1 (transformation mark corresponding to $Id=R_0$);
   \item chord 3 to chord 1: harmonic suavity = tonal suavity of chord $\VMreste_\text{chord 3}=\text{chord 1}$ 
	with respect to the bass of chord 3 (i.e. replacing $f_0$ by 140.625).
\end{itemize}
It is important not to forget, as aforementionned, that $o_n=0$ when calculating
harmonic suavity using tonal suavity.

Harmonic suavity of the signal is equal to 69.92\% with Euler's decomposition, and to
89.43\% with the proposed one.

\medskip
Finally, the noise emitted by this  asymmetrical two-stroke engine as a
rhythmic suavity equals to 75\%, a tonal suavity equals to 36.02\% and
an harmonic suavity equals to 89.43\%.
Global suavity is equals to 42.10\%.

\subsection{\label{sec:lomb} Symmetrical two-stroke engine}

In this second example, the signal is the noise emitted by a symmetrical two-stroke engine.
One explosion happens at each crankshaft rotation.

The analysis of cycles yields a mean duration equal to 0.0551~s 
with a standard deviation of $1.10.10^{-3}$~s.
One cycle (which means 1 period, and also 1 crankshaft rotation this time)
corresponds to an engine rotation speed equals to 1089~rpm.

\begin{figure}
   \centerline{\epsfig{figure=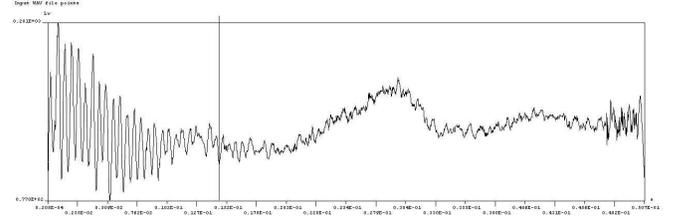,width=\columnwidth}}
   \caption{\label{fig:lombpulses}Detected pulses}
\end{figure}

Pulse analysis on this cycle leads to acoustic events as depicted by vertical
lines in figure Fig.~\ref{fig:lombpulses}.
Only 2 acoustic events are detected at points 1 and 871 in a cycle containing 2433 points.
This correspond to 2 acoustic events at pulses 0 and 2 within a cycle containing a total of 5 pulses.
The rhythmic suavity is equal to $2/5=0.40$.

\medskip
Tonal centre is chosen as the fundamental frequency $f_0=11.7$~Hz, and the tonal suavity 
is equal to 37.85\% with the proposed decomposition.

\medskip
The analysis of chords transitions is reduced to transitions from chord 1 to chord 2 and from
chord 2 to chord 1.
In both cases, we have $\VMdihedral=R_{10}\circ S_0$.
Harmonic suavity is equal to 70.00\% in both cases.

\medskip
Finaly, global suavity is equals to 23.21\%.

\medskip
Comparison between asymmetrical and symmetrical two-stroke engines yields that:
\begin{itemize}
   \item Contrary to intuition, the asymmetrical engine has a better rhythmic suavity than the 
	symmetrical engine. This point is confirmed by a jury.
   \item tonal suavities are almost the same for both engines. This means that chords of both engines 
	has the same overall beauty.
   \item The difference is more pronounced on the harmonic suavity.
	This means that the transition from a chord to the other is smoother for the asymmetrical engine.
\end{itemize}
A way to improve the noise of the symmetrical engine (compared to the asymmetrical engine) 
could be to add one frequency to the second acoustical event (in order to improve the harmonic 
suavity, but without lowering the tonal suavity), and to work on the rythmicity of the engine (for
example by adding a third acoustic event).

\section{\label{sec:ccl} Conclusion}

The purpose of the present study was to develop a tool able to give a measure 
of the beauty, or the acoustical quality, of (periodic) noises.

To perform this task, we decomposed this beauty into three components:
rhythmic, tonal and harmonic suavity, each of them giving an indication
of beauty according to a different point of view.

From an algorithmic point of view, suavity computations are performed by
manipulating only array of integers, which consumes little memory and 
requires very low computation time.

\medskip
It may be interesting to notice that harmonic suavity is a way of coding a chord
from the previous one by only the dihedral and the non congruous
components. 

\medskip
Although we have so far used a $\Z/12\Z$ decomposition, it is quite possible to use
any other decomposition the moment an easyness was formalized (i.e. as in tables~\ref{tab-euler} 
and \ref{tab-VMt}):
for example the tha\"i-khmer musical scale uses $\Z/7\Z$, and the slendro scale used in Java 
uses $\Z/5\Z$\cite{bib-universalis}.

\medskip
Finally the presented method shows excellent agreement with the feeling of a human jury 
(8 internal members plus customers teams) on all industrial cases treated untill now.


\bibliography{VM_suavity}


\end{document}